# A General Architecture for Behavior Modeling of Nonlinear Power Amplifier using Deep Convolutional Neural Network

Xin Hu[1,*], Zhijun Liu[1,*], You Li[2], Lexi Xu[3], Sun Zhang[1], Qinlong Li[4], Jia Hu[5], Wenhua Chen[6], Weidong Wang[1], Mohamed Helaoui[2] & Fadhel M. Ghannouchi[2]

Nonlinearity of power amplifier is one of the major limitations to the achievable capacity in wireless transmission systems. Nonlinear impairments are determined by the nonlinear distortions of the power amplifier and modulator imperfections. The Volterra model, several compact Volterra models and neural network models to establish a nonlinear model of power amplifier have all been demonstrated. However, the computational cost of these models increases and their implementation demands more signal processing resources as the signal bandwidth gets wider or the number of carrier aggregation. A completely different approach uses deep convolutional neural network to learn from the training data to figure out the nonlinear distortion. In this work, a low complexity, general architecture based on the deep real-valued convolutional neural network (DRVCNN) is proposed to build the nonlinear behavior of the power amplifier. With each of the multiple inputs equivalent to an input vector, the DRVCNN tensor weights are constructed from training data thanks to the current and historical envelope-dependent terms, I, and Q, which are components of the input. The effectiveness of the general framework in modeling single-carrier and multi-carrier power amplifiers is verified.

1 School of electronic engineering, Beijing University of Posts and Telecommunications, Beijing 100876, China. 2 Schulish School of Engineering, University of Calgary, Calgary T2N 1N4, Canada. 3 Network Technology Research Institute, China United Network Communications Corporation, Beijing 100048, China. 4 Electromagnetics and communication laboratory, Xi'an Jiaotong University, Xi'an 710049, China. 5 Computer Science, University of Exeter, Exeter EX4 4QJ, UK. 6 Department of Electronic Engineering, Tsinghua University, Beijing 100084, China. * These authors contribute equally to this work. Correspondence and requests for materials should be addressed to X. H. and Z. J. L. (email: huxin2016@bupt.edu.cn, lzj2017110489@bupt.edu.cn).

Nonlinear distortions caused by semiconductor components are commonly observed in a wide range of applications spanning from low-frequency power electronics to radio-frequency (RF) electronics [1]–[3]. In communication applications, the power amplifier (PA) is a key element that influences the overall performances of the transmitter in terms of both signal fidelity (i.e., linearity) and power efficiency [4], [5]. In the presence of wideband signals having non-constant amplitudes, the PA behaves as a dynamic nonlinear system that exhibits static distortions and memory effects [6] – [8]. Accordingly, the Volterra series appeared as the most comprehensive behavioral model. Since it is impractical to employ full Volterra series due to the large number of coefficients they include, several compact versions have been proposed. These can be classified in two categories, i.e., pruned Volterra series [9], [10] and memory polynomial (MP)-based models [11]. MP models are widely used in behavioral modeling of transmitters and PAs exhibiting memory effects as they achieve a tradeoff between computational cost and accuracy. In addition to PA nonlinearity, modulator imperfections as detailed in [12] and [13], also deteriorate the performance of the wireless transmitting system and degrade the quality of the output signal. Many methods have been proposed to consider the issues of in-phase and quadrature components (I/Q) imbalance and dc offset [14], [15].

Recently, a different approach was taken using machine learning algorithms [16]-[18]. These algorithms represented by the neural network (NN) aim to construct nonlinear impairments by learning from data. The NN model not only exhibits a strong fitting ability to the nonlinearity of PA, but also has an ideal performance in the presence of various distortions. Due to their strong non-linear approximation ability, NN has a great attraction to build the behavioral modeling of PAs. Techniques, such as a fuzzy logic NN [16], a feed-forward NN [17], [18], and a radial basis NN [19], have recently been suggested for constructing the nonlinear behavior of a device under test (DUT). However, the computational cost of these models increases and their implementation demands more signal processing resources as the signal bandwidth gets wider or the number of carrier aggregation. This motivates the development of computationally efficient NN-based behavioral models.

In this paper, it is shown that an enhanced model based on a deep real-valued convolutional neural network (DRVCNN) in which a pre-designed filter based on deep convolutional computation model is proposed to learn the features of input signals, uses the tensor representation model to extend the CNN from the vector space to the tensor space. The tensor weights are used to build the model with an extra degree of generality and flexibility, which is demonstrated to achieve an accurate model at a complexity lower than other models. Furthermore, due to the introduce of the attention module, DRVCNN not only learns from the training data and generates a model of the power amplifier, but also guides us how to reduce the complexity by distinguishing the terms. Another advantage of the proposed method is that each carrier can be represented by a vector so that it can be adapted to PA modeling for different types of communication signals. Since the model based on DRVCNN becomes free from specifics of the PA, it can be applied universally to all PA modeling whether they are narrowband PA, wideband PA, single-band PA, multi-band PA, or whether they are MIMO systems or the state-of the art. It is also shown that this model is versatile and robust enough that the training can be performed at the transmitter. The pre-training of the pre-designed filter further reduces the training cost of the model. Experiments on 100M single signal, 40MHz dual carrier signal and 20MHz triple-carrier signal verify the effectiveness of the proposed method. Compared with the traditional classic generalized memory polynomial (GMP) model, the normalized mean square error (NMSE) performance is improved by about 2dB; compared with the NN-based ARVTDNN model, the complexity of the proposed model is reduced by about 50% and the NMSE performance is not lost.

## Results

**Input features:** In order to learn the dynamic characteristics of nonlinear power amplifier, a DRVCNN model is proposed. Considering the nonlinearity and memory effects of the transmitter, the I/Q components and envelope dependent terms of the current and past signals of each carrier signal need to be included in the input data. We define the input matrix formed by the $k$-th carrier signal as

$$X_k(n) = \big[ I_k(n), I_k(n-1),...,I_k(n-M); \\ Q_k(n), Q_k(n-1),...,Q_k(n-M); \\ |x_k(n)|, |x_k(n-1)|,...,|x_k(n-M)|; \\ |x_k(n)|^2, |x_k(n-1)|^2,...,|x_k(n-M)|^2; \\ |x_k(n)|^3, |x_k(n-1)|^3,...,|x_k(n-M)|^3 \big] \quad (1)$$

where $X_k(n)$ is a matrix of $5 \times (M+1)$; $I_k(n)$ and $Q_k(n)$ represent I/Q components of the $k$-th carrier signal $x_k(n)$, respectively; $|x_k(n)|$ denotes the amplitude of the $k$-th carrier signal $x_k(n)$; $I_k(n-i)$, $Q_k(n-i)$ and $|x_k(n-i)|, (i=1,2,\cdots,M)$ denote the corresponding terms of past samples, respectively; $M$ represents the memory depth.

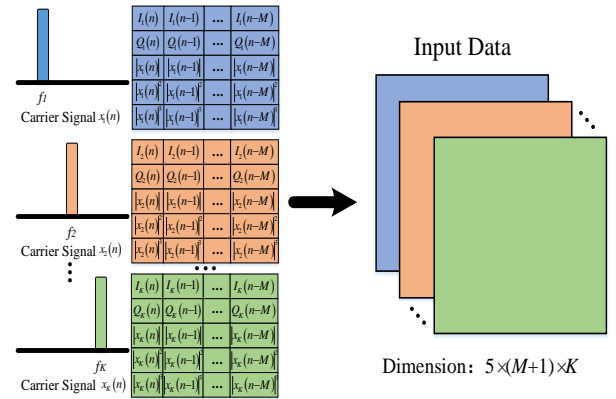

**Fig. 1** Structure diagram of input data.

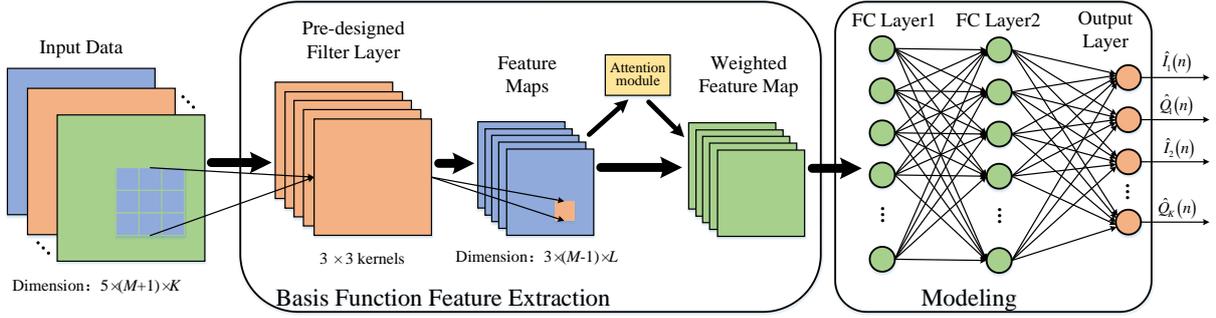

**Fig. 2** Block diagram of proposed DRVCNN model.

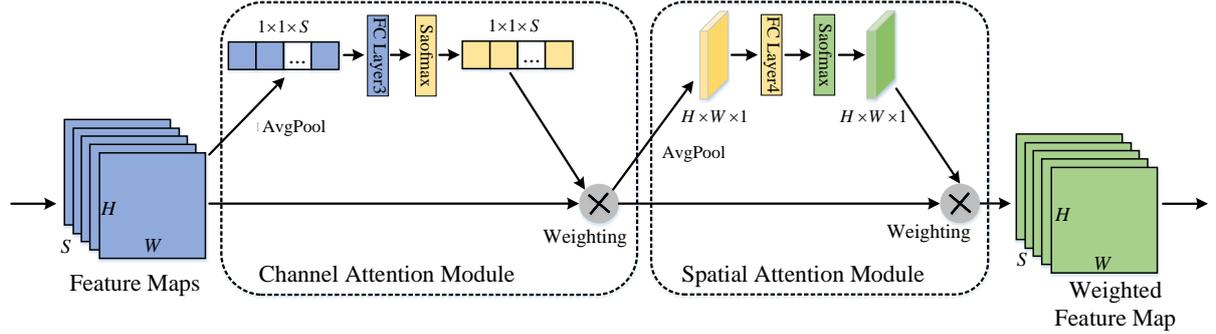

**Fig. 3** Design block diagram of the attention module.

To consider the intermodulation between different carrier signals, the input data needs to include the input matrix of all carrier signals. In order to reduce the complexity of the model, superimposing the input matrix of different carrier signals into a three-dimensional tensor is considered. The input data of the model is expressed as

$$\mathbf{X}_n = [X_1(n), X_2(n), ..., X_K(n)] \quad (2)$$

where $\mathbf{X}_n$ is a tensor of $5 \times (M+1) \times K$, $K$ is number of carrier signals.

**Structure of the DRVCNN model:** The DRVCNN model contains a pre-designed filter, an attention module, two fully connected layers and an output layer, as shown in Fig. 2. The pre-designed filters consist of a number of pre-trained convolutional layer and each pre-designed filter contains $3K$ convolution kernels with dimensions of 3*3. The convolution operation of the pre-designed filter is represented as

$$\mathbf{R} = f_c(\mathbf{X}_n \otimes \boldsymbol{\omega}_c + \mathbf{b}_c) \quad (3)$$

where $\mathbf{R}$ are the feature maps of the output of the pre-designed filter, and the dimension is $H \times W \times S$, and $S$ is the number of convolution kernels; $\boldsymbol{\omega}_c$, $\mathbf{b}_c$ are convolution kernels and biases, respectively; $f_c(\cdot)$ is the activation function, which is set to 'tanh'.

To improve the modeling performance of the model, after the pre-designed filters, an attention module is introduced to emphasize the important feature points in the feature map. The traditional attention module requires weighting of all feature points, which will cause unbearable complexity. The literature [20] proposes to weight feature points separately from the channel and spatial dimensions to reduce the implementation complexity of the attention module. This paper uses this structure to implement the attention module, as shown in Fig. 3. The feature map undergoes an average pooling operation from $H$ and $W$ to obtain a map with dimensions $1 \times 1 \times S$. Then, a fully connected layer, which includes $S/3$ neurons, 'Tanh' activation function and a Softmax layer with $S$ neurons, integrate the features of the map to generate a weight matrix $\mathbf{W}_a$. The feature map weighted from the channel dimension is expressed as

$$\mathbf{R}_a = \mathbf{R} \times \mathbf{W}_a \quad (4)$$

In the spatial attention module, the weighted feature map is weighted again from the spatial dimension. The feature graph $\mathbf{R}_a$ is subjected to an average pooling operation from the $S$ dimension to obtain a map with dimensions $H \times W \times 1$. Then, a fully connected layer, which includes $HW/3$ neurons, 'Tanh' activation function and a Softmax layer with $HW$ neurons, are used to generate the weight matrix $\mathbf{W}_s$. The final weighted feature maps are expressed as

$$\mathbf{R}_s = \mathbf{R}_a \times \mathbf{W}_s \quad (4)$$

After the attention module, two fully connected (FC) layers with activation function 'Tanh' are used to fit the PA output signal corresponding to each carrier. The number of neurons in the two FC layers is set to 5K and 3K, respectively. The proposed method is divided into two stages: training and execution stages. In the training stage, the DRVCNN learns from the training data and generates a model of the PA. In the execution stage, the nonlinear distortion is calculated based on the model, and the distortion can be removed based on the model.

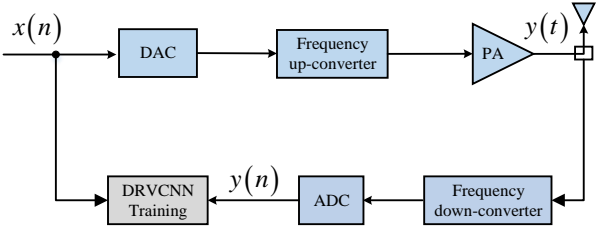

**Fig. 4** Modeling process of the nonlinear transmitter.

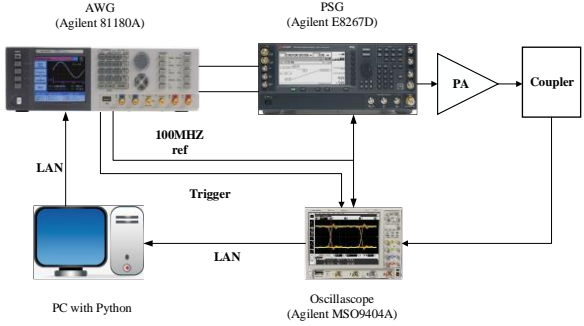

**Fig. 5** Experimental setup.

**Training stage:** To improve the energy efficiency of the transmitter, the PA needs to work close to the saturation point. The realization process of nonlinear PA modeling is shown in Fig. 4. The baseband modulated signal is feed into the digital-to-analog conversion (DAC) module and the up-conversion module. Then, a coupler is used to capture the output of the transmitted signal. After the down-conversion module and the analog-to-digital conversion (ADC) module, the baseband signal is captured. Using the baseband modulated signal and the sampled baseband signal, the modeling of the nonlinear PA can be build. As the training does not have to operate at the data rate unlike the execution stage, and the amount of training data is not excessive, it can be performed offline to save computation cost. To test the proposed DRVCNN model, an experimental platform was built, as shown in Fig. 5. The test signals were a 100MHz single carrier orthogonal frequency division multiplexing (OFDM) signal, a 40MHz dual-carrier OFDM signal and a 28MHz three carrier OFDM signal. The signal generator E8267D implements DAC and frequency-up conversion. The test device is a GaN Doherty PA with a small-signal gain (SSG) of 13 dB and a center frequency of 2.14 GHz, and its saturation power is 43 dBm. The Keysight 89600 Vector Signal Analyzer (VSA) software running on the oscilloscope (MSO) 9404A implements the frequency-down conversion and ADC. Three uncorrelated datasets with different types of signals are generated for training and testing. Each pair of signals contains 20K input and output sample samples.

The modeling data is divided into a training set and a test set according to the ratio of 3:2, which are used for model training and testing respectively. The loss function of the model is defined as the mean square error (MSE) function. The training of the proposed model is divided into two steps: the training of the pre-designed filters and the modeling of the PA. First, the parameters of the DRVCNN model are updated according to the Adam optimization algorithm [21] using training data. The trained convolutional layer is regarded as a pre-designed filter bank and fixed. Then, based on the pre-designed filters with fixed parameters, the PA is modeled by adjusting the attention module and the fully connected layer using the Adam optimization algorithm. The modeling performance is described by normalized mean square error (NMSE).

$$NMSE_i = 10 \times \lg \frac{\frac{1}{N}\sum_{n=1}^{N}\left(\left(\hat{I}_i(n)-I_{outi}(n)\right)^2+\left(\hat{Q}_i(n)-Q_{outi}(n)\right)^2\right)}{\frac{1}{N}\sum_{n=1}^{N}\left(\left(I_{outi}(n)\right)^2+\left(Q_{outi}(n)\right)^2\right)} \quad (5)$$

where $NMSE_i$ is the NMSE performance of the $i$-th carrier signal; $\hat{I}_i(n)$ and $\hat{Q}_i(n)$ are the predicted I/Q components of the $i$-th carrier signal, respectively; $I_{outi}(n)$ and $Q_{outi}(n)$ are the measured I/Q components of the $i$-th carrier signal, respectively. The training process of the RVTDCNN model is shown in algorithm 1.

---

**Algorithm 1** Training of the proposed model

**Definition:**
1. Define the input structure according to the number of carrier signals and the structure of the DRVCNN model;
2. Obtain training data containing input data and label data;
3. Define the loss function as the MSE function;

**Extraction of parameters of the Pre-designed Filter:**
1. Training the DRVCNN model:
  **Loop**: l=1, 2, …, L1
    1) Calculate the output of the model and the MSE function;
    2) Judgment: if MSE requirements are met, exit the entire loop;
    3) Update model parameters according to Adam algorithm;
  End
2. Save the parameters of the convolutional layer and define them as pre-designed filter parameters.

**PA Modeling:**
Training the DRVCNN model:
  **Loop**: l=1, 2, …, L2
    1) Calculate the output of the pre-designed filter using fixed pre-designed filter parameters;
    2) Calculate model output and loss function;
    3) Judgment: if MSE requirements are met, exit the entire loop;
    4) Update model parameters according to Adam algorithm;
**End**

---

**Execution stage.** As shown in Fig. 2, the DRVCNN model is implemented on a 100MHz single carrier signal. The NMSE performance expressed in Eq. (5) is used to evaluate the model performance. The optimal network structure of the proposed DRVCNN model and other typical methods under 100MHz single carrier OFDM signal is shown in Table 1.

**Table 1** Network structure of different methods (*M*=3).

| Parameter | Setting |
|---|---|
| **DRVCNN** | |
| Input data dimension | 5*4 |
| Num. of convolution kernels | 3 |
| Size of convolution kernels | 3*3 |
| FC layer structure | [5 3] |
| Number of neurons in FC3 | 1 |
| Number of neurons in FC4 | 2 |
| Num. of output neuron | 2 |
| **ARVTDNN in [17]** | |
| Input data dimension | 20 |
| Num. of neurons in FC layer | 17 |
| Activation (FC layer) | 'Tanh' |
| Num. of output neuron | 2 |
| **DNN [18]** | |
| Input data dimension | 8 |
| Hidden layer structure | [17 17 17] |
| Activation (hidden layer) | 'Tanh' |
| Num. of output neuron | 2 |
| **GMP in [10]** | |
| The index arrays for aligned signal and envelope | $K_a$=11, $L_a$=7 |
| The index arrays for signal and lagging envelope | $K_b$=3, $M_b$=2, $L_b$=5 |
| index arrays for signal and leading envelope | $K_c$=0, $M_c$=0, $L_c$=0 |

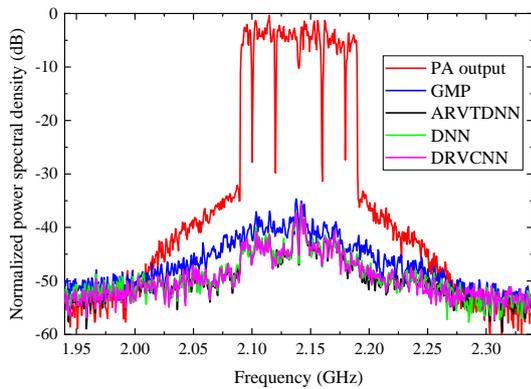

**Fig. 6** Spectral comparison of modeling errors between DRVCNN and other models at 100 MHz single carrier OFDM source signal.

**Table 2** Comparison of modeling performance and complexity between the proposed method and other methods.

| | Num. of model coefficients | NMSE (dB) |
|---|---|---|
| GMP | 214 | -33.35 |
| ARVTDNN | 393 | -36.65 |
| DNN | 801 | -36.52 |
| DRVCNN | 193 | -36.73 |

Fig. 6 shows the modeling error spectrum of the proposed model and other methods under a 100MHz single carrier source signal. The modeling error in the figure represents the difference between the measured output of the transmitter and the predicted output of these models. It can be found that the modeling error spectrum of DRVCNN is much smaller than that of the GMP model regardless of the transmission channel or the adjacent channel, indicating the superiority of the modeling performance of the proposed DRVCNN model. At the same time, the modeling error spectrum of the DRVCNN model is almost equal to that of the DNN model and ARVTDNN model, indicating that the modeling performance of these methods is almost equal.

Table 2 compares the model complexity and modeling performance (NMSE performance) of DRVCNN model with other typical models. In the table, the model complexity is expressed by the number of model coefficients. It can be found that compared with the traditional GMP model, the NMSE performance of the DRVCNN model can be improved by about 3dB, while the complexity is also reduced. This is because the traditional GMP model cannot improve the modeling performance by adding polynomial bases in the broadband case due to the high correlation between polynomial bases. Compared with ARVTDNN model and DNN model, the number of model coefficients of DRVCNN model is reduced by more than 50%, and the modeling performance is not reduced.

To further verify the modeling performance of the DRVCNN model for transmitters under different carrier numbers, the proposed DRVCNN model in Fig. 2 is implemented on a 40MHz dual-carrier signal. The modeling performance of each carrier signal of the model is evaluated by the NMSE performance expressed by Eq. (5). The optimal network structure of the proposed DRVCNN model and other typical methods under 40MHz dual-carrier OFDM signal is shown in Table 3.

**Table 3** Network structure of different methods ($M$=3).

| Parameter | Setting |
|---|---|
| **DRVCNN** | |
| Input data dimension | 5*4*2 |
| Num. of convolution kernels | 6 |
| Size of convolution kernels | 3*3 |
| FC layer structure | [10 6] |
| Number of neurons in FC3 | 2 |
| Number of neurons in FC4 | 2 |
| Num. of output neuron | 4 |
| **ARVTDNN in [17]** | |
| Input data dimension | 40 |
| Num. of neurons in FC layer | 35 |
| Activation (FC layer) | 'Tanh' |
| Num. of output neuron | 4 |
| **DNN [18]** | |
| Input data dimension | 16 |
| Hidden layer structure | [25 25 25] |
| Activation (hidden layer) | 'Tanh' |
| Num. of output neuron | 4 |
| **GMP in [10]** | |
| The index arrays for aligned signal and envelope | $K_a$=9, $L_a$=5 |
| The index arrays for signal and lagging envelope | $K_b$=3, $M_b$=2, $L_b$=3 |
| index arrays for signal and leading envelope | $K_c$=0, $M_c$=0, $L_c$=0 |

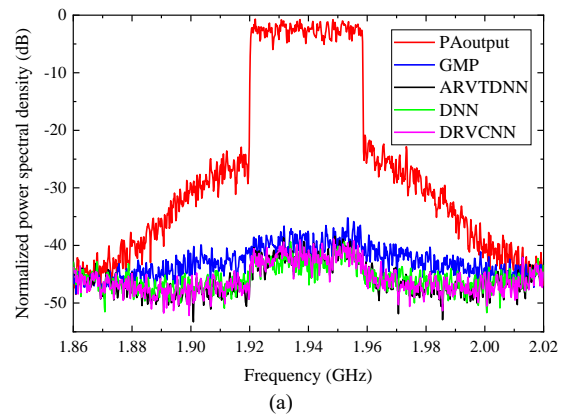

(a)

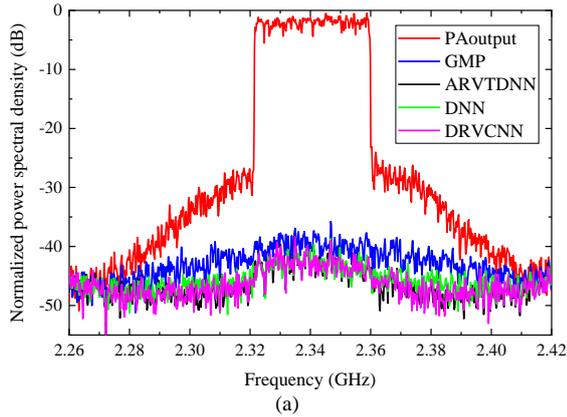

**Fig. 7** Spectral comparison of modeling errors between DRVCNN and other models at 40 MHz dual carrier OFDM source signal. (a) The first carrier signal; (b) The second carrier signal

**Table 4** Comparison of modeling performance and complexity between the proposed method and other methods in dual-carrier source signals.

|  | Num. of model coefficients | NMSE (dB) |
|---|---|---|
| GMP | 1400 | -33.82/-34.19 |
| ARVTDNN | 1579 | -36.62/-37.45 |
| DNN | 1829 | -36.01/-36.46 |
| DRVCNN | 636 | -36.65/-37.67 |

Fig. 7 shows the spectrum of the modeling error of the proposed model and other methods under the 40MHz dual-carrier source signal. It can be found that the error spectrum of the two carrier signals of the proposed method is less than -40dB for both the transmit channel and the adjacent channel, indicating the effectiveness of the proposed method for modeling dual-carrier transmitters. Compared with the GMP model, the DRVCNN model has more superior modeling performance, which can be seen from the error spectrum. Meanwhile, the modeling error spectrum of the DRVCNN model is almost equal to that of the ARVTDNN model and DNN model, indicating that the modeling performance of these methods is almost equal.

Table 4 compares the model complexity and modeling performance (NMSE performance) of the DRVCNN model with other typical models under the 40MHz dual-carrier source signal. The results show that the NMSE performance of the proposed DRVCNN model can be improved by about 3dB on two carrier signals compared to the traditional GMP model, while the model complexity is reduced by about 50%. Compared with the ARVTDNN model, the number of model coefficients of the DRVCNN model has been reduced by about 60%, while the modeling performance has not deteriorated. Compared with the DNN model, the number of model coefficients of the proposed model has been reduced by about 65%, and the modeling performance is also improved.

The proposed DRVCNN model in Fig. 2 is implemented on a 20MHz triple-carrier signal. The modeling performance of each carrier signal of the model is evaluated by the NMSE performance expressed by Eq. (5). The optimal network structure of the proposed DRVCNN model and other typical methods under 20MHz triple-carrier OFDM signal is shown in Table 5.

**Table 5** Network structure of different methods ($M=2$).

| Parameter | Setting |
|---|---|
| **DRVCNN** |  |
| Input data dimension | 5*3*3 |
| Num. of convolution kernels | 9 |
| Size of convolution kernels | 3*3 |
| FC layer structure | [15 9] |
| Number of neurons in FC3 | 3 |
| Number of neurons in FC4 | 1 |
| Num. of output neuron | 6 |
| **ARVTDNN in [17]** |  |
| Input data dimension | 45 |
| Num. of neurons in FC layer | 40 |
| Activation (FC layer) | 'Tanh' |
| Num. of output neuron | 6 |
| **DNN [18]** |  |
| Input data dimension | 18 |
| Hidden layer structure | [30 30 30] |
| Activation (hidden layer) | 'Tanh' |
| Num. of output neuron | 6 |
| **GMP in [10]** |  |
| The index arrays for aligned signal and envelope | $K_a=7$, $L_a=4$ |
| The index arrays for signal and lagging envelope | $K_b=3$, $M_b=1$, $L_b=2$ |
| index arrays for signal and leading envelope | $K_c=0$, $M_c=0$, $L_c=0$ |

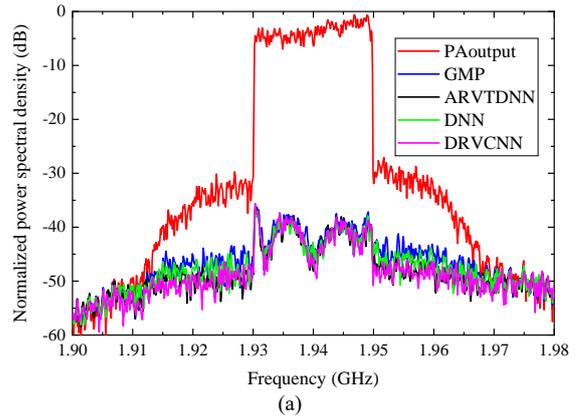

(a)

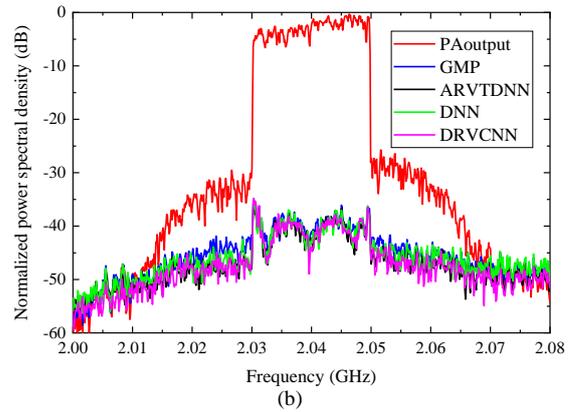

(b)

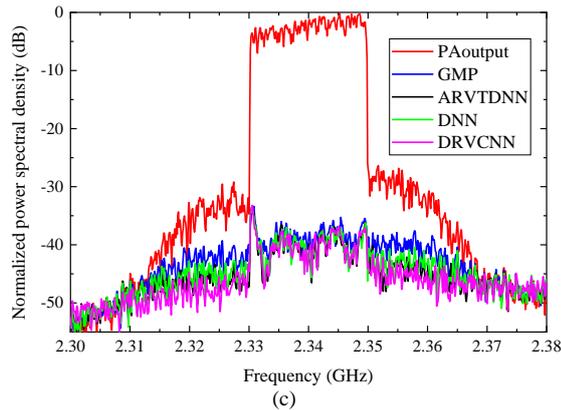

(c)

**Fig. 8** Spectral comparison of modeling errors between DRVCNN and other models at 20 MHz triple-carrier OFDM source signal. (a) The first carrier signal; (b) The second carrier signal; (c) Third carrier signal.

**Table 6** Comparison of modeling performance and complexity between the proposed method and other methods in triple-carrier source signals.

|  | Num. of model coefficients | NMSE (dB) |
|---|---|---|
| GMP | 2880 | -33.89/-32.91/-32.80 |
| ARVTDNN | 2086 | -36.50/-35.37/-35.07 |
| DNN | 2616 | -35.74/-34.34/-34.33 |
| DRVCNN | 943 | -36.49/-35.43/-35.11 |

Fig. 8 shows the spectrum of modeling errors of the proposed DRVCNN model and other methods under a 20MHz triple-carrier source signal. The results show that the error spectrum of the DRVCNN model is less than -40dB regardless of the transmission channel or the adjacent channel, which shows the effectiveness of the method for modeling triple-carrier transmitters. At the same time, the DRVCNN model has a smaller modeling error than the GMP model, and the modeling performance is better. Compared with the ARVTDNN model, the modeling performance of the DRVCNN model has not decreased.

Table 6 compares the model complexity and modeling performance (NMSE performance) of the DRVCNN model with other typical models under the 20MHz triple-carrier source signal. The results show that, compared with the traditional GMP model, the NMSE performance of the DRVCNN model on three carrier signals can also reduce the NMSE by about 2dB, and the model complexity is reduced by about 67%. Compared with the ARVTDNN model and the DNN model, the number of model coefficients of the DRVCNN model has been reduced by more than 57%, while the modeling performance has not deteriorated.

# References

1. John, L., Tessmann, A., Leuther, A., Neininger, P., Merkle, T., Zwick, T. Broadband 300-GHz Power Amplifier MMICs in InGaAs mHEMT Technology. *J. IEEE Trans. THz Sci. Technol.* **10**, 309-320 (2020).
2. Dyck, A. et al. A transmitter system-in-package at 300 GHz with an off-chip antenna and GaAs-based MMICs. *J. IEEE Trans. THz Sci. Technol.* **9**, 335–344 (2019).
3. Diebold, S. et al. A novel 1 ×4 coupler for compact and high-gain power amplifier MMICs around 250 GHz. *J. IEEE Trans. Microw. Theory Tech.* **63**, 999–1006 (2015).
4. Mosleh, S. S., Liu, L., Sahin, C., Zheng, Y. R., Yi, Y. Brain-Inspired Wireless Communications: Where Reservoir Computing Meets MIMO-OFDM. *J. IEEE Trans. Neural Netw. Learn. Syst.* **29**, 4694-4708 (2018).
5. Larose, C. L., Ghannouchi, F. M. Optimal adaptation methods and class of operation: keys to improving feedforward amplifier power efficiency. *J. EEE Trans. Veh. Technol.* **54**, 456-467 (2005).
6. Reina-Tosina, J., Allegue-Mart ńez, M., Crespo-Cadenas, C., Yu, C., Cruces, S. Behavioral modeling and predistortion of power amplifiers under sparsity hypothesis. *J. IEEE Trans. Microw. Theory Techn.* **63**, 745–753 (2015).
7. Zhang, Q., Chen, W., Feng, Z. Reduced cost digital predistortion only with in-phase feedback signal. *J. IEEE Microw. Wireless Compon. Lett.* **28**, 257–259 (2018).
8. Joung, J., Ho, C. K., Adachi, K., Sun, S. A survey on poweramplifier-centric techniques for spectrum-and energy-efficient wireless communications. *J. IEEE Commun. Surv. Tut.* **17**, 315–333 (2015).
9. Braithwaite, R. N. Digital Predistortion of an RF Power Amplifier Using a Reduced Volterra Series Model with a Memory Polynomial Estimator. *J. IEEE Trans. Microw. Theory Techn.* **65**, 3613–3623 (2017).
10. Morgan, D. R., Ma, Z., Kim, J., Zierdt, M. G., Pastalan, J. A generalized memory polynomial model for digital predistortion of RF power amplifiers. *J. IEEE Trans. Signal Process.* **54**, 3852–3860, (2006).
11. Belabad, A. R., Motamedi, S.A. A novel generalized parallel two-box structure for behavior modeling and digital predistortion of RF power amplifiers at LTE applications. *J. Circuits Syst. Signal Process.* **37**, 2714–2735 (2018).
12. Rawat, M., Ghannouchi, F. M. A mutual distortion and impairment compensator for wideband direct-conversion transmitters using neural networks. *J. IEEE Trans. Broadcast.* **58**, 168–177 (2012).
13. Lajnef, S., Boulejfen, N., Abdelhafiz, A., Ghannouchi, F. M. Two dimensional Cartesian memory polynomial model for nonlinearity and I/Q imperfection compensation in concurrent dual-band transmitters. *J. IEEE Trans. Circuits Syst. II, Exp. Brief* **63**, 14–18 (2016).
14. Cavers, J. K. New methods for adaptation of quadrature modulators and demodulators in amplifier linearization circuits. *J. IEEE Trans. Veh. Technol.* **46**, 707–716 (1997).
15. Kim, Y.-D., Jeong, E.-R., Lee, Y. H. Adaptive compensation for power amplifier nonlinearity in the presence of quadrature modulation/demodulation errors. *J. IEEE Trans. Signal Process* **55**, 4717–4721 (2007).
16. Chen, C. L. P., Wang, J., Wang, C.-H., Chen, L. A new learning algorithm for a fully connected neuro-fuzzy inference system. *J. IEEE Trans. Neural Netw. Learn. Syst.* **25**, 1741–1757 (2014).
17. Wang, D., Aziz, M., Helaoui, M., Ghannouchi, F. M. Augmented real-valued time-delay neural network for compensation of distortions and impairments in wireless transmitters. *J. IEEE Trans. Neural Netw. Learn. Syst.* **30**, 242–254 (2019).
18. Hongyo, R., Egashira, Y., Hone, T. M., Yamaguchi, K. Deep Neural Network-Based Digital Predistorter for Doherty Power Amplifiers. *J. IEEE Microw. Wireless Compon. Lett.* **29**, 146–148 (2019).
19. Robnik-Šikonja, M. Data generators for learning systems based on RBF networks. *J. IEEE Trans. Neural Netw. Learn. Syst.* **27**, 926–938 (2016).
20. Woo, S., Park, J., Lee, J. Y., Kweon, I. S. CBAM: Convolutional Block Attention Module. In Proc. *2018 European Conference on Computer Vision (ECCV 2018)*, p.1-17 (Munich, Germany,2018).
21. Kingma D. P., Ba, J. Adam: A Method for Stochastic Optimization. In *Proc. 3rd International Conference for Learning Representations (ICLR 2015)*, p.1-3 (San Diego, USA, 2015).